\documentclass[final,conference,twocolumn]{IEEEtran}

\usepackage{graphicx}
\usepackage{cite}
\usepackage{amsthm}
\usepackage{amsmath,amssymb,amsfonts}
\usepackage{mathtools}
\usepackage{bm}
\usepackage{enumitem} 
\usepackage{cases}
\usepackage{empheq}
\usepackage{xcolor}
\usepackage{acro}
\usepackage{mathtools}

\DeclareAcronym{AWGN}{short = AWGN ,long = additive white Gaussian noise}
\DeclareAcronym{ACRDA}{short = ACRDA ,long = asynchronous contention resolution diversity ALOHA}
\DeclareAcronym{CDF}{short = CDF ,long = cumulative distribution function}
\DeclareAcronym{CRA-CC}{short = CRA-CC ,long = CRA-convolutional code}
\DeclareAcronym{CRA-SH}{short = CRA-SH ,long = CRA-shannon bound}
\DeclareAcronym{CRA}{short = CRA ,long = contention resolution ALOHA}
\DeclareAcronym{CRDSA}{short = CRDSA ,long = contention resolution diversity slotted ALOHA}
\DeclareAcronym{CRDSA++}{short = CRDSA++ ,long = contention resolution diversity slotted ALOHA++}
\DeclareAcronym{CRI}{short = CRI ,long = contention resolution interval}
\DeclareAcronym{CSA}{short = CSA ,long = coded slotted ALOHA}
\DeclareAcronym{CSI}{short = CSI ,long = channel state information}
\DeclareAcronym{DAMA}{short = DAMA ,long = demand assigned multiple access}
\DeclareAcronym{DSA}{short = DSA ,long = diversity slotted ALOHA}
\DeclareAcronym{DSSS}{short = DSSS ,long = direct sequence spread spectrum}
\DeclareAcronym{E_SSA}{short = E-SSA ,long = enhanced spread spectrum ALOHA}
\DeclareAcronym{ECRA}{short = ECRA ,long = enhanced contention resolution ALOHA}
\DeclareAcronym{ECRA-SC}{short = ECRA-SC ,long = ECRA selection combining}
\DeclareAcronym{ECRA-MRC}{short = ECRA-MRC ,long = ECRA maximal-ratio combining}
\DeclareAcronym{EGC}{short = EGC ,long = equal-gain combining}
\DeclareAcronym{FEC}{short = FEC ,long = forward error correction}
\DeclareAcronym{GEO}{short = GEO ,long = geostationary orbit}
\DeclareAcronym{HAP}{short = HAP ,long = high-altitude platform}
\DeclareAcronym{IC}{short = IC ,long = interference cancellation}
\DeclareAcronym{IoT}{short = IoT ,long = Internet of things}
\DeclareAcronym{IRA}{short = IRA ,long = irregular repetition ALOHA}
\DeclareAcronym{IRCRA}{short = IRCRA ,long = irregular repetition contention resolution ALOHA}
\DeclareAcronym{IRSA}{short = IRSA ,long = Irregular repetition slotted ALOHA}
\DeclareAcronym{LDPC}{short = LDPC ,long = low-density parity-check}
\DeclareAcronym{LEO}{short = LEO ,long = low-Earth orbit}
\DeclareAcronym{M2M}{short = M2M ,long = machine-to-machine}
\DeclareAcronym{MAC}{short = MAC ,long = medium access control}
\DeclareAcronym{MF}{short = MF ,long = matched filter}
\DeclareAcronym{MF-TDMA}{short = MF-TDMA ,long = multi-frequency time division multiple access}
\DeclareAcronym{MRC}{short = MRC ,long = maximal-ratio combining}
\DeclareAcronym{MUD}{short = MUD ,long = multiuser detection}
\DeclareAcronym{pmf}{short = pmf ,long = probability mass function}

\DeclareAcronym{PDF}{short = PDF ,long = probability density function}
\DeclareAcronym{PER}{short = PER ,long = packet error rate}
\DeclareAcronym{PLR}{short = PLR ,long = packet loss rate}
\DeclareAcronym{QPSK}{short = QPSK ,long = quadrature phase-shift keying}
\DeclareAcronym{RA}{short = RA ,long = random access}
\DeclareAcronym{RCB}{short = RCB ,long = random coding bound}
\DeclareAcronym{RTT}{short = RTT ,long = round trip time}
\DeclareAcronym{SA}{short = SA , long = slotted ALOHA}
\DeclareAcronym{SB}{short = SB ,long = Shannon bound}
\DeclareAcronym{SC}{short = SC ,long = selection combining}
\DeclareAcronym{SIC}{short = SIC ,long = successive interference cancellation}
\DeclareAcronym{SNIR}{short = SNIR ,long = signal-to-noise and interference ratio}
\DeclareAcronym{SINR}{short = SINR ,long = signal-to-interference and noise ratio}
\DeclareAcronym{SNR}{short = SNR ,long = signal-to-noise ratio}
\DeclareAcronym{TDMA}{short = TDMA ,long = time division multiple access}
\DeclareAcronym{UCP}{short = $\Code$-UCP ,long = $\Code$-unresolvable collision pattern}
\DeclareAcronym{VF}{short = VF ,long = virtual frame}

\DeclareAcronym{rv}{short = r.v., long = random variable}

\usepackage{balance}

\mathtoolsset{showonlyrefs}
\newtheorem{example}{Example}

\IEEEoverridecommandlockouts

\begin{document}
 

 \title{Two-Step Interference Cancellation for Energy Saving in Irregular Repetition Slotted ALOHA}
 \author{

\IEEEauthorblockN{Estefan\'{\i}a Recayte\IEEEauthorrefmark{1},  Leonardo Badia\IEEEauthorrefmark{2}, and Andrea Munari\IEEEauthorrefmark{1}}   
\IEEEauthorblockA{\IEEEauthorrefmark{1} Institute of Communications and Navigation, German Aerospace Center (DLR), Germany } 
\IEEEauthorblockA{\IEEEauthorrefmark{2} Dept. Information Engineering, University of Padova, Italy}
\IEEEauthorblockA{email: estefania.recayte@dlr.de, leonardo.badia@unipd.it, andrea.munari@dlr.de}

  \thanks{\small E. Recayte and A. Munari acknowledge  the financial support by the Federal Ministry of Education and Research of Germany in the programme of “Souverän. Digital. Vernetzt.” Joint project 6G-RIC, project identification number: 16KISK022.}
 }

 \maketitle
 


\thispagestyle{empty} \pagestyle{empty}

\begin{abstract}
We evaluate a modification of irregular repetition slotted ALOHA (IRSA) involving intermediate decoding and early transmission termination by some nodes, upon their decoding success. This is meant to avoid unnecessary transmissions, thereby reducing energy consumption. We expect this to be particularly useful at low loads, where most transmissions can be avoided as they do not often result in a collision and are therefore redundant.
To validate this proposal, we observe that most of the literature related to IRSA considers an asymptotic heavily loaded regime; thus, we also present a model of energy consumption and success probability for frames of limited length and low offered loads. Thanks to our analysis, also confirmed by simulation, we are able to show that the proposed technique is able to reduce IRSA energy consumption by minimizing transmissions, while preserving performance gains over standard ALOHA. For example, we are able to get a $33\%$ energy saving at offered loads around $10\%$ without affecting throughput.\end{abstract}



\newcommand{\users}{\mathsf{m}}
\newcommand{\slots}{\mathsf{n}}
\newcommand{\load}{\mathsf{G}}
\newcommand{\userDegree}{r}
\newcommand{\avgUser}{\bar{\mathsf{r}}}
\newcommand{\slotDegree}{l}
\newcommand{\area}[1]{a_{#1}}
\newcommand{\SNIR}{\mathsf{\Gamma}}
\newcommand{\x}{\mathsf{x}}
\renewcommand{\epsilon}{\varepsilon}

\newcommand{\puserToslot}{\mathsf{p}}
\newcommand{\qslotTouser}{\mathsf{q}}

\newcommand{\estef}{\textcolor{blue}}
\newcommand{\change}{\textcolor{orange}}

\newcommand{\pmf}{\Lambda}
\newcommand{\prob}{p}
\newcommand{\energy}{\mathsf{E}}

\newcommand{\setS}{s}

\newcommand{\lo}{\mathsf{G}}
\newcommand{\rate}{\mathsf{R}}
\newcommand{\dt}{\gamma}

\newcommand{\ec}{\mathsf{E}}
\newcommand{\E}{\mathsf{E}}
\newcommand{\plr}{\mathtt{PLR}}
\newcommand{\tr}{\mathtt{T}}
\newcommand{\se}{\mathtt{S}}
\newcommand{\per}{\mathsf{P_{L}}}
\newcommand{\pdec}{\mathsf{P_{dec}}}

\newcommand{\ee}{\eta}

\newcommand{\nsl}{\mathsf{m}}

\newcommand{\numBit}{k}
\newcommand{\nsy}{\mathsf{L}}
\newcommand{\syd}{\mathsf{T_s}}

\newcommand{\noise}{\mathsf{N}}
\newcommand{\noiseVec}{\bm{\noise}}
\newcommand{\noiseVar}{\sigma_{\noise}^2}

\newcommand{\Pw}{\mathsf{p}}
\newcommand{\packetT}{\mathsf{t}}
\newcommand{\Int}{\mathsf{I}}
\newcommand{\Ns}{\mathsf{N}}
\newcommand{\fpw}{\alpha}
\newcommand{\nil}{\tau}

\newcommand{\sinr}{\gamma}

\newcommand{\mutInf}{\mathsf{I}}
\newcommand{\avMutInf}{\bar{\mutInf}}
\newcommand{\rvDec}{\mathcal{D}}

\newcommand{\us}{\mathsf{u}}
\newcommand{\rep}{\mathsf{r}}
\newcommand{\nin}{\mathsf{t}}

\newcommand{\nurv}{U}
\newcommand{\nus}{u}


\section{Introduction}\label{sec:introd}


 \ac{IRSA} \cite{Liva:IRSA} is an enhanced version of framed slotted ALOHA (FSA), which relies on a bipartite graph optimization of contention resolution. The rationale behind this methodology is to organize access in synchronous frames, comprising a known number of slots. During each frame, users select a number of slots, according to a certain distribution, and resulting in an irregular bipartite graph for user/slot association. The selected slots are independently occupied by the users with a replica of their packets. 

If the overall occupancy of one slot, considering the superposition of all sources, ends up as consisting of only one replica, the corresponding packet is successfully received. In addition, packets can be further decoded by leveraging coding and \ac{SIC}, applied exploiting the underlying structure of the bipartite graph, akin to iteratively-decodable error correction through message-passing \cite{KrishnaIRSA}.

IRSA and other coded slotted ALOHA techniques find application in massive access for the Internet of things (IoT), satellite communications, and wireless sensors networks, to improve the efficiency of communication, i.e., enhancing throughput and/or supporting a large number of devices to access the channel \cite{badia2021game, Enrico_IRSA}.

The focus of investigations pertaining to this and similar massive access techniques is often on highlighting how the bottleneck of traditional ALOHA procedures can be avoided \cite{nisioti2019fast,badia2024game}. While this is a required preliminary step, pervasive communication in the IoT, especially involving sensors with limited capabilities and/or located in remote areas, are also subject to strong needs in terms of \emph{energy efficiency} \cite{EEIRSA2022,badia2022discounted, faddoul2024spatial }.

The traditional justifications behind the efforts to limit the energy expenditure of the nodes lie in the difficulties of battery replacement \cite{tian2023centralized}, or the fact that communication often resorts to some form of energy harvesting, which may be scarce at times \cite{ma2019sensing}. This problem is even more acute in extreme scenarios like satellite links or emergency communications \cite{zhen2020energy,do2021joint}.

The goal of the present paper is to identify practical techniques to reduce the consumption of IRSA at the sender side, which ultimately boils down to diminishing the number of transmissions, without significantly affecting the resulting performance gain brought over the standard ALOHA access.
In particular, the main idea can be seen as inspired by detection techniques that allow for an early stop of useless transmissions. For example, in CSMA/CD, devices are able to detect collisions in real-time and stop transmitting when one happens \cite{xu2022design}. This helps to avoid wasting resources on transmitting corrupted data. 

Our rationale is similar, but based on detecting \emph{successful transmissions}, as opposed to failures, ahead of time. In IRSA, users with successfully decoded packets could stop transmitting further replicas before the end of the frame without any consequence. The irregular graph is just pruned of unnecessary arcs, since the further replicas of such nodes would be removed through \ac{SIC} \cite{toni2015prioritized}. The reduced number of transmissions will allow the involved nodes to save energy \cite{da2023energy}.

Such a shortcut cannot be adopted on a per-slot basis, as the successful decoding of replicas requires exploration of the irregular graph and message passing. Yet, an intermediate check can be adopted. We will therefore consider an IRSA variant where packet decoding is attempted half-way through the frame and the explained procedure of stopping transmission from users whose packets have been successfully decoded is applied only once. Multiple subdivision are of course possible but with a tradeoff between the increased control and the actual energy saving. Repeating the procedure in each slot will cause the system to become a plain slotted ALOHA.

In the existing literature, most studies consider saturation of asymptotically long frames \cite{IRSAasympHa,SICShieh}. In our case, instead, finite frame lengths have to be considered, since we need to include the impact of splitting SIC over subsequent sub-frames. In such settings, important contributions were obtained \cite{AlexEF}, yet focusing on reliability and packet loss rate. This prompts us to circumvent most limitations on SIC, but at the same time we derive analytical and approximated expressions of energy consumption for low channel loads.

Our finding, verified by both analysis and simulation, is that the proposed technique allows for a reduced number of transmissions and therefore decreased consumption, which is particular relevant at low channel loads. For instance, we show that for a commonly studied distribution, more than $30\%$ energy saving is obtained for loads around $0.1$ [packets/slot].

The rest of this paper is organized as follows. Sect.~\ref{sec:sysmodel} presents the system model, while in Sect.~\ref{sec:AEC} the average energy consumption  is derived. In Sect.~\ref{sec:approx}, an approximate consumption model for low to moderate loads is proposed. Numerical results are given in Sect.~\ref{sec:results}, and Sect.~\ref{sec:Conclusions} concludes.

\vspace{0.15cm}
\noindent {\it Notation:} 
We use capital letters, e.g. $X$, for \acp{rv} and their lower case counterparts, e.g. $x$, for their realizations. The probability mass function (pmf) of a discrete \ac{rv} $X$ is denoted as ${\Pr\{X=x\} = P_X(x)}$. Furthermore, we denote conditional pmfs as ${\Pr\{X=x | Y = y\} = P_{X|Y}(x|y)}$ and the expectation of $X$ as $ \mathbb{E}[X] = \sum_x x P_X(x)$.


\vspace{0.15cm}
\section{System Model}\label{sec:sysmodel} 

We consider a system with $\users$   uncoordinated active nodes (\emph{users}), which transmit packets to a common receiver by sharing a wireless channel. Time is divided into slots of equal duration, during the transmission of a single packet. A \ac{MAC} frame consists of $\slots$ consecutive slots, and users are both slot and frame-synchronized. The channel load, denoted with $\load$, is defined as
\begin{equation}\label{eq:load}
  \load = \frac{\users}{\slots} \quad [\text{packets/slots}].
\end{equation}

\begin{figure}[t]
  \includegraphics[width=\columnwidth]{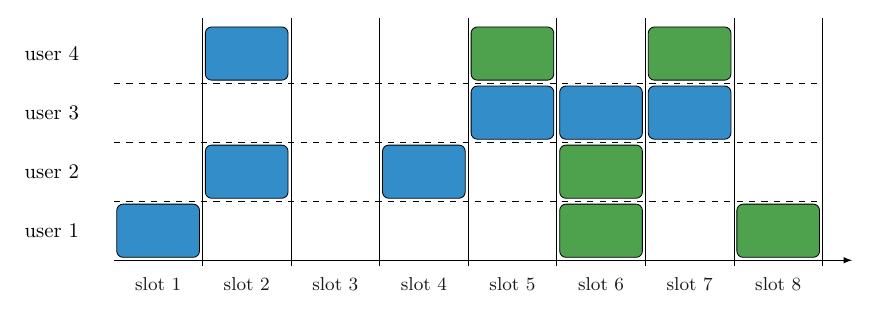}
     \centering \caption{Transmission pattern over a \ac{MAC} frame of $\slots=8$ slots and with $\users=4$ active users for IRSA distribution   $\Lambda(\x) = \x^3$. }
     \label{fig:sysmodel}
  \end{figure}

Following the \ac{IRSA} protocol \cite{Liva:IRSA}, each node shall send $r$ copies (or \emph{replicas}) of its data packet uniformly, placed over the $\slots$ available slots.
In practice, $r$ is independently drawn by each user according the \ac{IRSA} probability distribution, and $r \in \{2, r_{\max} \} $. Following the traditional notation \cite{PaoliniGraph}, we can write the polynomial formulation of the IRSA distribution as
\begin{equation}\label{eq:IRSAdist}
  \Lambda(\x) = \sum_{r=2}^{r_{\max}} \Lambda_{r} \,\x^r
\end{equation}
where $\Lambda_r$ denotes the probability that a node transmits $r$ replicas. Each replica contains a pointer to the slot indexes where the other replicas are located.\footnote{This can be obtained, e.g., by placing the pointers into the packet header. A more efficient approach uses the packet payload as the seed of the random generator used by the sender to place its replicas. Thus, upon decoding one of the replicas, the receiver can retrieve the positions of all other ones.} This information is used by the receiver during the decoding process. As example of the frame structure in IRSA with $ \Lambda(\x) =  \x^3$ is shown in Fig.~\ref{fig:sysmodel}, considering $\users = 4$ users, and $\slots =8$ slots.


Denoting the \ac{rv} associated to the number of replicas that a user plans to transmit in the entire frame as $R$, the average number of replicas transmitted per user is 
\[  \mathbb{E}[R] = \sum_{r=2}^{r_{\max}} \Lambda_r \, r. \] 
Throughout our discussion, we assume collisions as destructive, so that the receiver cannot directly extract information from a slot where multiple users transmitted. Conversely, a slot occupied by a sole packet (singleton slot), leads to decoding.

At the receiver's side, \ac{SIC} is performed. 
The algorithm operates by iteratively decoding replicas that are interference free. When a replica is decoded, the receiver learns where the other replicas of the same packet are located, and their contribution  from the incoming signal is removed. Ideal interference cancellation is assumed. As a consequence, replicas from other users may become interference free and can thus be decoded. This is repeated until no interference free replicas are identified or up to a predefined number of iterations \cite{EF_Petar}.

\subsection*{Two-step SIC approach}
To improve  the energy efficiency of \ac{IRSA}, we consider the following modification to the scheme. Every node continues to draw the number of replicas at the start of the frame. However, only replicas falling under the first $\alpha \leq \slots$ are sent at first. We refer to the initial $\alpha$ slots as the first part (or portion) of the frame. 
At the receiver's side, a first decoding attempt is performed after $\alpha$ slots applying the \ac{SIC} algorithm to the first portion of the frame, as described earlier. After this procedure, the receiver distributes a feedback to all users informing them of which packets (i.e., nodes) have been successfully retrieved. 
We observe that the proposed solution entails an extra costs due to the transmission of the feedback after $\alpha$ slots. This causes an additional energy expenditure for the devices, which have to receive the broadcast message. Nonetheless, as highlighted by the results in Section~\ref{sec:results}, this cost is possibly outweighed by the energy savings. 

Along this line, we note that the feedback can be provided as a short message. For instance, one viable technique is to transmit a packet of length $\alpha$ bits, in which the $i$-th bit is set to one if packets in the $i$-th slot were successfully decoded, and zero otherwise.  
By examining the relevant positions within the feedback packet, each node is able to determine whether or not its replicas have been successfully decoded. 

Then, the second part of the SIC approach takes place. In this case, nodes that were already decoded no longer send others replicas, whereas uncoded nodes continue operating as originally planned and sending further replicas in the remaining $\slots -\alpha$ slots. At the end of the frame, the decoder process all $\slots$ slots through the SIC algorithm.

To illustrate this approach, 
let us consider again the transmission model illustrated in Fig.~\ref*{fig:sysmodel}, where we assume that the SIC processes $\alpha = 4$ slots at the first step. At the first step, user  1 is decoded in slot 1 as it is a singleton slot. Following this, user 2 is decoded in slot 4, and its copy from slot 2 is cancelled. Subsequently, user 4 becomes free from interference and is also successfully decoded. The receiver informs these three users to cease their transmissions, and replicas colored in green are spared from further transmission.

\section{Average Energy Consumption}\label{sec:AEC}
\begin{figure*}[t!]
      \setcounter{equation}{8}
   \begin{align}
   \begin{split}
       \E =  \sum_{r=2}^{r_{\max}}   \Lambda_r  \, \Big\{ \sum_{t=0}^r \sum_{h=2}^{r_{\max}} \Lambda_h \binom{h}{t} \prod_{i=0}^{t-1} \frac{\alpha -i}{\slots -i}   \prod_{j=0}^{h - t -1 } \frac{\slots - \alpha -j}{\slots - t- j}\big[t + (r- t) \, (1- \pdec(t))  \big] \Big\} \,  \Pw \,   \packetT.
   \end{split}
   \label{eq:Energyfinal}
    \end{align}
   \hrulefill
\end{figure*}

We denote by $\Pw$  the transmission power, set as the same value for all nodes, and by $\packetT$ the packet duration. The energy consumed per packet transmission is given by 
\begin{equation} 
      \setcounter{equation}{1}
\label{eq:energy}
  \mathsf{e} = \Pw \cdot \packetT \quad \text{[joule].}
\end{equation}
Let $K$  be the \ac{rv} associated to the number of replicas that a node transmits in the two-step SIC scheme. 
Note that $K=R$ if the node is not decoded during the first step of the SIC, and $K \leq R$ otherwise.
We aim to evaluate the average energy consumed per node by the proposed scheme, which can be expressed as
\begin{equation}\label{eq:energytot}
    \E = \mathbb{E}[K] \cdot \mathsf{e} \leq \mathbb{E}[R] \cdot \mathsf{e}.
\end{equation}

The average number of replicas transmitted per user  depends on the behavior of the system in the first  $\alpha$ slots. Specifically, we need to calculate the distribution of the replicas transmitted by a user in the first part of the frame, and the probability  of such user to be resolved through \ac{SIC} at the first step.

We derive the replica distribution for a user in $\alpha$ slots as follows.
Take \ac{rv} $T$ as the number of replicas sent.
Given the initial IRSA replica distribution $\Lambda_r$, and assuming $\alpha \geq r_{\max}$, the user can transmit $t \in [0,1, \dots , r_{\max}]$ replicas in $\alpha$ slots. Leaning on the polynomial notation, the replica probability distribution $\Gamma(\x)$ in the first portion of the frame is

\begin{equation}
    \Gamma(\x) = \sum_{t=0}^{r_{\max}}  \Gamma_t \, \x^t 
\end{equation}
where $\Gamma_t = P_T(t)$ is derived as follows
\begin{align}\label{eq:lambdat}
    \Gamma_t & = \sum_{r=2}^{r_{\max}} P_{T|R}(t|r)P_R(r) \\
    &  = \sum_{r=2}^{r_{\max}} \Lambda_r \binom{r}{t} \prod_{i=0}^{t-1} \frac{\alpha -i}{\slots -i}   \prod_{j=0}^{r - t -1 } \frac{\slots - \alpha -j}{\slots - t- j}
 \end{align}
 where $\Lambda_r = P_R(r)$. The binomial coefficient accounts for all possible ways of choosing $t$ replicas from the $r$ to transmit. In turn, the first product represents the probability that $t$ replicas are placed in the first part of the frame, while the second product  the probability that $r-t$ replicas are placed at the end of the frame.  In the ratio, the denominator represents all potential slots to place a replica,
 while the numerator indicates the number of available slots to position the replica in the part of the frame under consideration (positive cases).
 
 For better comprehension, evaluate this numerical example.
 \begin{example}\label{ex:ex1}
   Assuming a frame length of $\slots = 200 $ and the distribution given by $ \Lambda(\x) =  \x^3$, consider a SIC over the first half of the frame, i.e. ${\alpha = 100}$ slots. The evaluation of the probabilities leads to  ${\Gamma_0  =0.1231},  {\Gamma_1  = 0.3769}, \Gamma_2 = 0.3769 $ and $ \Gamma_3 = 0.1231 $.  The polynomial distribution becomes
   \begin{equation}
   \Gamma(\x) =   0.1231 \,  \x^0 + 0.3769 \, \x^1 +  0.3769 \, \x^2 + 0.1231 \, \x^3.
   \label{eq:gammgamm}
   \end{equation}
   Note that when only a portion of the frame is taken into account, the replica distribution also includes strictly positive probabilities for users placing none or one of its replicas  in the first part of the frame, i.e. $\Gamma_0$ and $\Gamma_1$. 
   In this numerical example, $\Gamma_0 = \Gamma_3$ and  $ \Gamma_1 =  \Gamma_2$ due to the fact that $\alpha = \slots /2$.
   Another particular case is when $\alpha = \slots$, which clearly collapses into $\Gamma(\x) = \Lambda(\x)$.
 \end{example}

 
 The probability that a user transmits a total of $K$ replicas,  given that $R$  were selected from the original distribution and $T$ were sent at the beginning of the frame, is expressed as
 \begin{equation}\label{eq:pkrt}
    P_{K|R, T}(k|r,t) = \begin{cases}
        \pdec(t)      &\, \, \text{for } k = t  \\
        1 - \pdec(t)  & \text{ for }   k = r.
    \end{cases} 
 \end{equation}
 Here, $\pdec(t)$ denotes the probability that a user is decoded at the first step of the SIC process, having sent $T=t$ in the first $\alpha$ slots.
We have that 
\begin{align} 
    \mathbb{E}[K] & = \sum_k k \,  P_K(k) \\
     & = \sum_k k \, \sum_{r=2}^{r_{\max}} \, \sum_{t=0}^{r} P_{K|R, T}(k|r,t) P_{T|R}(t|r) P_R(r)
\end{align}
and, given \eqref{eq:pkrt}, we compute
\begin{align}\label{eq:avgK} 
        \mathbb{E}[K] & = \sum_{r=2}^{r_{\max}}   \Lambda_r  \, \Big\{ \sum_{t=0}^r \Gamma_t \big[t + (r- t) \, (1- \pdec(t)) \big] \Big\}.
 \end{align}
If we insert \eqref{eq:lambdat}, \eqref{eq:avgK}, and \eqref{eq:energy} into \eqref{eq:energytot}, we obtain  the expression of the average energy consumed per node given in  \eqref{eq:Energyfinal} (see top of next page). 
We further assess the successful delivery of packets versus the overall energy expenditure. To this end, we define the energy-normalized throughput $\eta$, as
\begin{equation}\label{eq:energyeff}
    \eta = \frac{\load [ 1- \mathsf{PLR(\load)}]}{\E} \quad \text{ [bit/joule]}
\end{equation}
where $\mathsf{PLR(\load)}$ is the packet loss rate indicating the probability that a packet is not successfully decoded in the frame, which is function of the channel load $\load$. 
We also define the packet loss when  a node transmits $t$ replicas  at the first part of the frame as  $\per(t) = 1 - \pdec(t).$
In the next section, we provide an approximation of the packet loss valid for low channel loads to evaluate \eqref{eq:avgK} and \eqref{eq:energyeff}.


\section{Packet Loss Approximation for Low- Moderate Channel Load} \label{sec:approx}
The  average energy and  energy efficiency evaluations presented earlier require computing the packet loss probabilities in the first part of the frame, $\per(t)$,  and overall,  $\mathsf{PLR(\load)}$.
Deriving the analytical expression for these quantities poses a challenge due to the  difficulty in   tracking the behavior of the \ac{SIC} process. Existing tools such as  density evolution can be implemented only assuming an infinite population of nodes placing replicas over a infinite frame, as in \cite{IRSAasympHa}. 

On the other hand, this technique cannot be applied in the case of finite length frames.
However, in the case of low to moderate channels load,   the packet loss of SIC can be approximated by addressing the so call \emph{stopping sets} \cite{AlexEF, EF_Petar,
EF_AlexG}.  These are configurations where multiple users transmit their replicas in the same time slots, resulting in a loop where the receiver cannot decode because none of the replicas are free from interference. 
A stopping set $\setS$ is defined by four 
elements
\begin{enumerate}[label=(\roman*)]
    \item  the number  $\omega(s)$ of users involved 
    \item the number $\mu(s)$ of slots involved
    \item the number $c(s)$ of isomorphisms, and 
    \item the set profile, denoted by $\boldsymbol{\nu}(s)$.
\end{enumerate}
The set profile of $s$ is represented as \[\boldsymbol\nu(s) = [\nu_0(s), \nu_1(s), ... ,\nu_{r_{\max}}(s)] \] 
where $\nu_y(s)$ is the number of users involved in the stopping set $s$ and transmitting  $y$ replicas. 
Given a set profile, the isomorphisms of a  stopping set are all the possible combinations of users and slots for which  the set profile can occur.
All of the aforementioned elements can be computed through enumeration.
    \begin{figure}[t]
    \hspace{-0.5cm}
    \includegraphics[width=1.044\columnwidth]{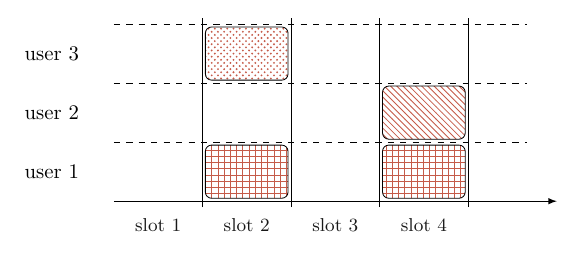}
       \centering \caption{Stopping set  example of three users transmitting their replicas in two slots. The stopping set profile is $\boldsymbol\nu(s)= [0,2,1,0]$. This represents a stuck situation in the two-step SIC approach.  }
       \label{fig:stopping_set}
    \end{figure}   
For a better understanding through examples, look at the following practical case.
\begin{example}
    Consider the stopping set depicted in Fig.~\ref{fig:stopping_set}. Here, user $1$ transmits two replicas in slots 2 and 4; user $2$ transmits one replica in slot 4, and user $3$ transmits one replica in slot 2. This represents a stopping set,  as the decoder fails to resolve any of these users. This stopping set can only occur in the two-step \ac{SIC} scheme when a portion of the frame is processed. In the traditional scheme, this particular stopping set does not occur, because users always transmit more than one replica, as indicated by distribution \eqref{eq:IRSAdist}. 
    In this stopping set, the number of users is $\omega(s) = 3$, the number of slots is $\mu(s)=2$, the set profile is  $\boldsymbol\nu(s) = [0,2,1,0]$, and the number of isomorphisms is $c(s) =2$.
    Indeed, another isomorphism corresponds to a transmission pattern where users $2$ and $3$ send replicas in slots $2$ and $3$, respectively. 
\end{example}
If we denote by $\mathcal{S}$ the ensemble of possible stopping sets that might occur when processing $\alpha$ slots, the probability that a users´employing the distribution $\Gamma(\x)$ is not decoded in the first part of the frame can be approximated in the low to moderate channel loads as in \cite{AlexEF}  

\begin{align}
\setcounter{equation}{9}
     \mathsf{P_a}(\alpha, \Gamma(\x))  \approx  \sum_{\setS \in \mathcal{S}} \phi(s) \omega(s ) c(s) \binom{\alpha}{\mu(s)}
     \prod_{t=1}^{r_{\max}} \frac{\Gamma_t^{\nu_h(s)}}{\nu_h(s)!}\binom{\alpha}{t}^ {\nu_t(\setS)} 
    \label{eq:Perrappox}
\end{align} 

\begin{flalign}
\textrm{where }    \phi(s)=\sum_{i=0}^{\omega(\setS )-1} (-1)^{\omega(\setS)-1+i } \frac{(\omega(\setS ) -1)!}{i!} (\slots \load)^i. &&
\nonumber
\end{flalign}
In general, an evaluation of \eqref{eq:energytot} requires the probability of not decoding a user conditioned on having sent exactly $T=t$ replicas in the first $\alpha$ slots. For simplicity, we approximate this quantity leaning on \eqref{eq:Perrappox}, i.e.,, on the average loss rate given the distribution $\Gamma(\x)$. We will then consider 
\[{\per(t) = 1 - \pdec(t) \simeq \mathsf{P_a}(\alpha, \Gamma(\x))}.\]
The tightness of this approximation will be verified in Sec.~\ref{sec:results}.

Note also that, for $\alpha = \slots$, which implies that $\Gamma(\x) =\Lambda(\x)$ then the probability that a user is not decoded in the the entire frame can be written as 
\begin{align}
\mathsf{PLR(\load)}=\mathsf{P_a}(\slots, \Lambda(\x)).
\end{align}

\section{Numerical Results and Discussion}\label{sec:results}
\begin{table}[t]
  \begin{center}
    \caption{Parameters of considered stopping sets}\label{tab:res}
    \begin{tabular}{r|c|c|r|c||r|c|c|r|c} 
  $s\!$      & $\boldsymbol\nu(s)$ & $\!\!\mu(s)\!\!\!$ & $\!\!c(s)\!\!\!$  & $\!\!\omega(s)\!\!\!$ & 
  $s\!$      & $\boldsymbol\nu(s)$ & $\!\!\mu(s)\!\!\!$ & $\!\!c(s)\!\!\!$  & $\!\!\omega(s)\!\!\!$ \\
      \hline
      \hline
     $1$   & $\!\![0,2,0,0]\!\!$ & $\!\!1\!\!$ & $1$  & $\!\!2$ &       
     $12\!$  & $\!\![0,2,0,2]\!\!$ & $\!\!4\!\!$ & $24$ & $\!\!4$\\
     $2$   & $\!\![0,0,2,0]\!\!$ & $\!\!2\!\!$ & $1$  & $\!\!2$&
     $13\!$  & $\!\![0,1,2,1]\!\!$ & $\!\!4\!\!$ & $24$ & $\!\!4$ \\
     $3$   & $\!\![0,2,1,0]\!\!$ & $\!\!2\!\!$ & $2$  & $\!\!3$&
     $14\!$  & $\!\![0,1,2,1]\!\!$ & $\!\!4\!\!$ & $24$ & $\!\!4$ \\
     $4$   & $\!\![0,0,0,2]\!\!$ & $\!\!3\!\!$ & $1$  & $\!\!2$&
     $15\!$  & $\!\![0,1,1,2]\!\!$ & $\!\!4\!\!$ & $48$ & $\!\!4$ \\
     $5$   & $\!\![0,1,1,1]\!\!$ & $\!\!3\!\!$ & $3$  & $\!\!3$&
     $16\!$  & $\!\![0,0,3,1]\!\!$ & $\!\!4\!\!$ & $24$ & $\!\!4$ \\
     $6$   & $\!\![0,0,3,0]\!\!$ & $\!\!3\!\!$ & $6$  & $\!\!3$&
     $17\!$  & $\!\![0,0,4,0]\!\!$ & $\!\!4\!\!$ & $72$ & $\!\!4$\\
     $7$   & $\!\![0,0,2,1]\!\!$ & $\!\!3\!\!$ & $6$  & $\!\!3$ &
     $18\!$  & $\!\![0,0,3,1]\!\!$ & $\!\!4\!\!$ & $\!\!144$ & $\!\!4$\\
     $8$   & $\!\![0,3,0,1]\!\!$ & $\!\!3\!\!$ & $6$  & $\!\!4$ &
     $19\!$  & $\!\![0,0,2,2]\!\!$ & $\!\!4\!\!$ & $48$ & $\!\!4$\\
     $9$   & $\!\![0,2,2,0]\!\!$ & $\!\!3\!\!$ & $\!\!12$ & $\!\!4$ &
     $20\!$  & $\!\![0,0,2,2]\!\!$ & $\!\!4\!\!$ & $48$ & $\!\!4$\\
     $10$  & $\!\![0,0,1,2]\!\!$ & $\!\!4\!\!$ & $\!\!12$ & $\!\!3$ &
     $21\!$  & $\!\![0,3,1,1]\!\!$ & $\!\!4\!\!$ & $72$ & $\!\!5$\\
      $11$  & $\!\![0,0,0,3]\!\!$ & $\!\!4\!\!$ & $\!\!24$ & $\!\!3$ &
      $22\!$  & $\!\![0,2,3,0]\!\!$ & $\!\!4\!\!$ & $\!\!144$ & $\!\!5$\\
          \end{tabular}
  \end{center}
\end{table}

In this section, numerical results obtained for the two-step SIC scheme are presented. The scheme is evaluated by means of simulations for low to high channel loads and analytically for low to moderate channel loads.

In all results, we consider frames of length $\slots =200$ slots and IRSA replica distribution within the whole frame given by ${\Lambda(\x) =  \x^3}$. We evaluate the two-step SIC process in two scenarios, i.e. $\alpha =  100$ slots and  $\alpha = 150$ slots. The case of $\alpha=100$ slots follow Example~\ref{ex:ex1} and thus its replica distribution in the initial portion of the frame is given in~\eqref{eq:gammgamm}, whereas for the case $\alpha = 150$ slots this distribution follows
\[\Gamma(\x) = 0.0149 \, \x^0 +  0.1399 \, \x^1 + 0.4254 \, \x^2 +  0.4198 \, \x^3.\]
We set both packet duration $\packetT$ and transmission power {$\Pw $} to be unitary. In this way, we focus on
the energy consumed per packet duration and neglect the unit of measure (joule).

\begin{figure}[t]
    \includegraphics[width=\columnwidth]{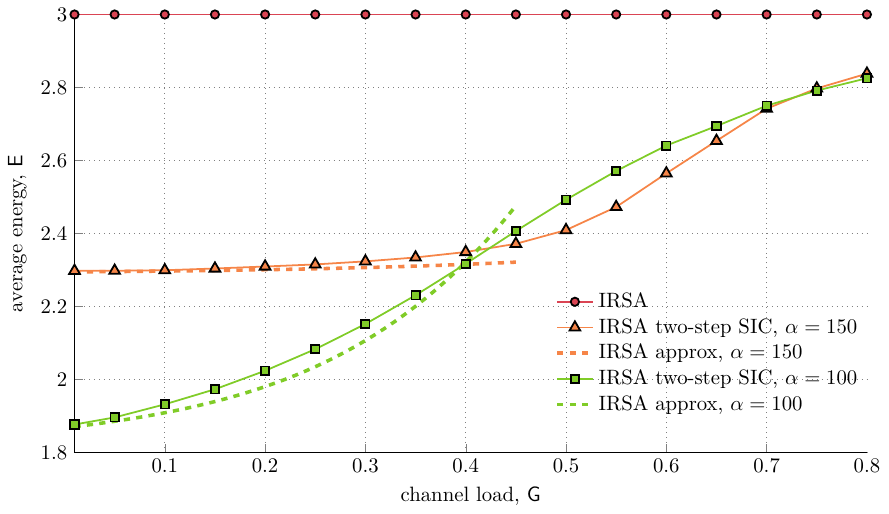}
       \centering \caption{Average energy consumption per node $\E$ as a function of channel load $\load$ for IRSA distribution $\Lambda(\x) = \x^3$, IRSA  with two-step SIC with $\alpha = 100$, and $\alpha =150$ slots. Curves with markers indicate  simulation results while dashed lines indicate  results provided by the analytical approximation. }
       \label{fig:Energy_avg}
    \end{figure}

We start evaluating the average energy consumption per user $\E$ as a function of the channel load $\load$ as shown in Fig.~\ref{fig:Energy_avg}. Curves with markers indicate results obtained by Monte Carlo simulations, while dashed curves results obtained through the proposed approximation. 
The stopping sets considered for the approximation of $\mathsf{P_a}(\alpha, \Gamma(\x))$ in \eqref{eq:Perrappox} are  all listed in Table~\ref{tab:res} together with the corresponding values of each parameter. 
    
In the IRSA scheme, since each user  always transmits three copies of its packet, the energy consumption per user is independent of the channel load and remains constant regardless of the value. A significant reduction in energy consumption is noted for the scheme utilizing the two-step SIC technique. 

In particular, the scheme in which the decoder processes the first $\alpha=150$ slots shows evident energy savings for each channel load value considered,  with a notable improvement of nearly $25\%$ observed for low channel loads.
For channel loads below $\load = 0.40$ [packets/slot],  we observe even higher savings when the decoder processes $\alpha = 100$ slots. In this case,  savings of over $33\%$ can be achieved with a simple modification to the protocol design.

It is worth noting that the energy consumption of the two-step SIC approach varies with the channel load.  This occurs since, as the number of users in the system increases, there is a higher probability that the decoder may not decode a user in the first part of the frame, due to more frequent occurrence of stopping sets. Failure to decode a user in the initial step of SIC requires transmitting additional replicas, consequently increasing the average consumption and approaching the performance of the basic IRSA operation. Concerning energy consumption, there exists a tradeoff between $\alpha$ and $\load$. Processing more slots in the initial step enhances the probability of decoding a user, as fewer stopping sets may occur.  However,  larger values of $\alpha$ also leads to sending more replicas.

Finally, the dashed curves confirm the accuracy of the analytical approximation for low and moderate loads in representing the real behavior of the proposed scheme, demonstrating its tightness.

    \begin{figure}[t]
    \includegraphics[width=\columnwidth]{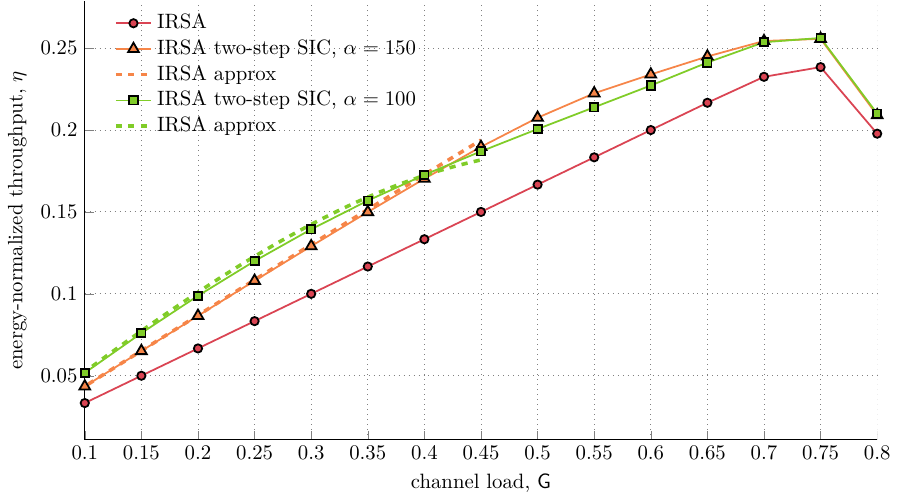}
       \centering \caption{ Energy-normalized throughput $\eta$ as a function of channel load $\load$ for IRSA distribution  $\Lambda(\x) = \x^3$, IRSA with two-step SIC with $\alpha = 100$, and $\alpha = 150$ slots.  Curves with markers indicate  simulation results while dashed lines indicate results provided by the analytical approximation. }
       \label{fig:energyeff}
    \end{figure}

In Fig.~\ref{fig:energyeff} the energy-normalized throughput $\eta$ as a function of the channel load $\load$ is plotted. Also for this picture, dashed curves show the proposed approximation and curves with markers report the simulation results.
For each scheme,  $\eta$ presents low values at low traffic due to the low throughput determined by $\load [ 1- \mathsf{PLR(\load)}]$.\footnote{We remark that the approximation of the packet loss rate $\mathsf{PLR(\load)}$ over the entire frame is computed through \eqref{eq:Perrappox} considering only stopping sets $s=4$ and $s=11$, i.e., the ones involving users sending three replicas. Indeed, when the whole frame is examined, the users which remain undecoded have transmitted all three replicas given that $\Lambda( \x)= \x^3$ is considered.  }
As expected the energy-normalized throughput for the two-step SIC scheme outperforms the traditional IRSA protocol for every channel load considered. Each scheme presents the maximum value of $\eta$ at the maximum throughput reached for the distribution $\Lambda(\x) = \x^3$ which in this case is around $\load = 0.75$ packets/slot.
This figure also demonstrates the validity of the given approximation by showing a tight behavior in response to the actual performance.

Additionally, we emphasize that the proposed two-step SIC preserves the performance in terms of packet loss rate, and consequently has the same throughput of the conventional approach. Indeed, it is easy to show that a node that is decoded using the IRSA protocol is also decoded using the two-step SIC, and vice versa for non solved nodes of IRSA.

\begin{figure}[t]
    \includegraphics[width=\columnwidth]{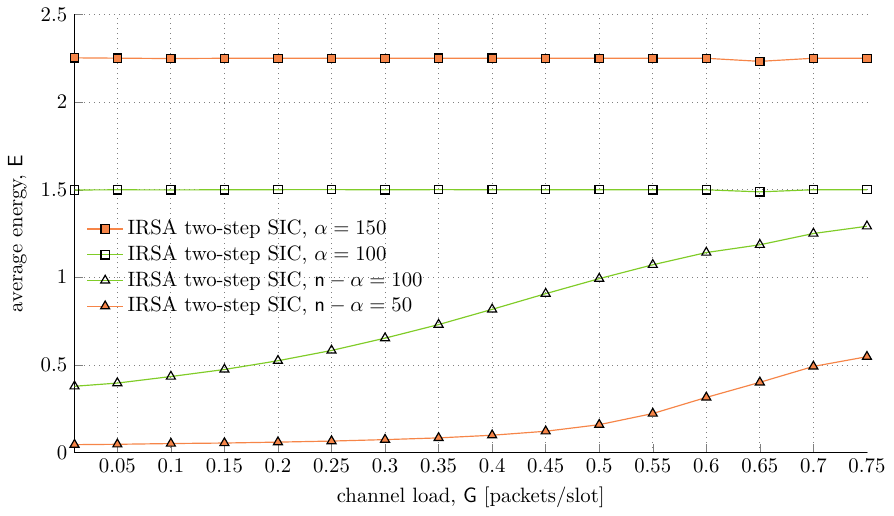}
       \centering \caption{Average energy consumption $\E$  per node as a function of channel load $\load$ for IRSA two-step at the first and second part of the frame.}
       \label{fig:Eperpart}
\end{figure}

Finally, in Fig.~\ref{fig:Eperpart} the average energy consumption $\E$  per node at each part of the frame for the two-step SIC as a function of the channel load is plotted.
Once again, we compare the cases where the first step considers of $\alpha=100$ or $\alpha=150$ slots, and we compare the consumption in the first and second part (square and triangle markers, respectively).
The energy consumption during the initial part of the frame is independent of the channel load considered. This happens because when $\alpha$ is fixed, the average number of replicas transmitted by a user within $\alpha$ slots remains constant as well. In this case, {$\E = 1.50$} per packet duration for $\alpha = 100$ while ${\E = 2.25 }$  for $\alpha = 150$.
As expected, the energy consumption in the second part of the frame increases when we  consider higher channel loads.  This behaviour arises because a more congested channel reduces the probability of successful decoding in the initial phase, requiring the transmission of additional replicas.

\section{Conclusions}\label{sec:Conclusions}

In this paper, we studied a two-step interference cancellation method for IRSA systems. Our research focuses on modification of the protocol to enhance overall system energy efficiency. We derived both close approximations and the exact formulations of the energy consumption. Our results confirm  notable improvements in energy saving, especially in systems operating at low to moderate channel load conditions. 
The proposed technique offers practical solutions with tangible benefits in real-world deployments contributing to the ongoing efforts to create more sustainable and efficient wireless communication networks.

\newpage
\bibliographystyle{IEEEtran}
\bibliography{IEEEabrv,references}

@STRING{IEEE_J_COML       = "{IEEE} Commun. Lett."}

@STRING{IEEE_J_JSAC       = "{IEEE} J. Sel. Areas Commun."}

@STRING{IEEE_J_COM        = "{IEEE} Trans. Commun."}

@STRING{IEEE_J_WCOM       = "{IEEE} Trans. Wireless Commun."}

@STRING{IEEE_J_WCOML      = "{IEEE} Wireless Commun. Lett."}

@STRING{IEEE_J_IT         = "{IEEE} Trans. Inf. Theory"}

@STRING{IEEE_J_IOT        = "{IEEE} Internet Things J."}

@STRING{IEEE_J_MC         = "{IEEE} Trans. Mobile Comput."}

@STRING{IEEE_J_CCN        = "{IEEE} Trans. on Cogn. Commun. Netw."}

@STRING{IEEE_O_CSTO       = "{IEEE} Commun. Surveys Tuts."}

@INPROCEEDINGS{KrishnaIRSA,
  author={Narayanan, Krishna R. and Pfister, Henry D.},
  booktitle={Proc. ISTC}, 
  title={Iterative collision resolution for slotted {ALOHA}: {A}n optimal uncoordinated transmission policy}, 
  year={2012},
  volume={},
  number={},
  pages={136-139},
  doi={10.1109/ISTC.2012.6325214}}

@ARTICLE{SICShieh,
  author={Shieh, Shin-Lin and Yang, Shih-Hung},
  journal=IEEE_J_COM, 
  title={{E}nhanced irregular repetition slotted {ALOHA} under {SIC} limitation}, 
  year={2022},
  volume={70},
  number={4},
  pages={2268-2280},
  doi={10.1109/TCOMM.2022.3147505}}

@ARTICLE{AlexEF,
  author={Sandgren, Erik and Graell i Amat, Alexandre and Brännström, Fredrik},
  journal=IEEE_J_COM, 
  title={{O}n frame asynchronous coded slotted {ALOHA}: {A}symptotic, finite length, and delay analysis}, 
  year={2017},
  volume={65},
  number={2},
  pages={691-704},
  doi={10.1109/TCOMM.2016.2633468}}

@ARTICLE{EF_Petar,
  author={Ivanov, Mikhail and Brännström, Fredrik and Graell i Amat, Alexandre and Popovski, Petar},
  journal=IEEE_J_COML, 
  title={Error Floor Analysis of Coded Slotted {ALOHA} Over Packet Erasure Channels}, 
  year={2015},
  volume={19},
  number={3},
  pages={419-422},
  doi={10.1109/LCOMM.2014.2385073}}

@INPROCEEDINGS{EF_AlexG,
  author={Paolini, Enrico and Liva, Gianluigi and Graell i Amat, Alexandre},
  booktitle={Proc. IEEE ICC}, 
  title={A structured irregular repetition slotted {ALOHA} scheme with low error floors}, 
  year={2017},
  volume={},
  number={},
  pages={},
  doi={10.1109/ICC.2017.7996564}}

@ARTICLE{Enrico_IRSA,
  author={Tralli, Velio and Paolini, Enrico},
  journal=IEEE_J_IT, 
  title={{IRSA}-based Random Access over the {Gaussian} Channel}, 
  year={2024},
  volume={},
  number={},
  doi={10.1109/TIT.2024.3353589}}

@ARTICLE{IRSAasympHa,
  author={Haghighat, Javad and Duman, Tolga M},
  journal=IEEE_J_WCOM, 
  title={An energy-efficient feedback-aided irregular repetition slotted {ALOHA} scheme and its asymptotic performance analysis}, 
  year={2023},
  volume={22},
  number={12},
  pages={9808-9820},
  doi={10.1109/TWC.2023.3273616}}

@ARTICLE{EEIRSA2022,
  author={Chen, Zhengchuan and Feng, Yifan and Tian, Zhong and Jia, Yunjian and Wang, Min and Quek, Tony Q. S.},
  journal=IEEE_J_WCOML, 
  title={Energy efficiency optimization for irregular repetition slotted {ALOHA}-based massive access}, 
  year={2022},
  volume={11},
  number={5},
  pages={982-986},
  doi={10.1109/LWC.2022.3151931}}

@ARTICLE{PaoliniGraph,
  author={Paolini, Enrico and Liva, Gianluigi and Chiani, Marco},
  journal=IEEE_J_IT, 
  title={{C}oded slotted {ALOHA}: {A} graph-based method for uncoordinated multiple access}, 
  year={2015},
  volume={61},
  number={12},
  pages={6815-6832},
  doi={10.1109/TIT.2015.2492579}}

@ARTICLE{Liva:IRSA,
  author={Liva, Gianluigi},
  journal= IEEE_J_COM, 
  title="Graph-based analysis and optimization of contention resolution diversity slotted  {ALOHA}", 
  year={2011},
  volume={59},
  number={2},
  pages={477-487},
  doi={10.1109/TCOMM.2010.120710.100054}}

@article{ma2019sensing,
  title={Sensing, {C}omputing, and {C}ommunications for {E}nergy {H}arvesting {IoTs}: {A} {S}urvey},
  author={Ma, Dong and Lan, Guohao and Hassan, Mahbub and Hu, Wen and Das, Sajal K},
  journal=IEEE_O_CSTO,
  volume={22},
  number={2},
  pages={1222--1250},
  year={2019},
}

@article{tian2023centralized,
  title={A centralized control-based clustering scheme for energy efficiency in underwater acoustic sensor networks},
  author={Tian, Wei and Zhao, Yangqing and Hou, Rui and Dong, Mianxiong and Ota, Kaoru and Zeng, Deze and Zhang, Junmim},
  journal={IEEE Trans. Green Commun. Netw.},
  year={2023},
volume = 7,
number = 2,
pages = {668--679},
}

@article{zhen2020energy,
  title={Energy-efficient random access for {LEO} satellite-assisted {6G} {Internet} of remote things},
  author={Zhen, Li and Bashir, Ali Kashif and Yu, Keping and Al-Otaibi, Yasser D and Foh, Chuan Heng and Xiao, Pei},
  journal=IEEE_J_IOT,
  volume={8},
  number={7},
  pages={5114--5128},
  year={2020},
}

@article{do2021joint,
  title={Joint optimisation of real-time deployment and resource allocation for {UAV}-aided disaster emergency communications},
  author={Do-Duy, Tan and Nguyen, Long D and Duong, Trung Q and Khosravirad, Saeed R and Claussen, Holger},
  journal=IEEE_J_JSAC,
  volume={39},
  number={11},
  pages={3411--3424},
  year={2021},
}

@article{xu2022design,
  title={Design and analysis of a novel collision notification scheme for {IoT} environments},
  author={Xu, Fangxin and Feng, Li and Yang, Jie and Liang, Hong and Chang, Yu-Teng},
  journal={J. Supercomp.},
  volume={78},
  number={16},
  pages={18130--18152},
  year={2022},
}

@inproceedings{badia2021game,
  title={A game theoretic approach to age of information in modern random access systems},
  author={Badia, Leonardo and Munari, Andrea},
  booktitle={Proc. IEEE Globecom Wkshps},
  year={2021},
}

@inproceedings{badia2022discounted,
  title={Discounted age of information for networks of constrained devices},
  author={Badia, Leonardo and Munari, Andrea},
  booktitle={Proc. IEEE MedComNet},
  pages={43--46},
  year={2022},
}

@article{toni2015prioritized,
  title={Prioritized random {MAC} optimization via graph-based analysis},
  author={Toni, Laura and Frossard, Pascal},
  journal=IEEE_J_COM,
  volume={63},
  number={12},
  pages={5002--5013},
  year={2015},
}

@article{nisioti2019fast,
  title={Fast {Q-learning} for improved finite length performance of irregular repetition slotted {ALOHA}},
  author={Nisioti, Eleni and Thomos, Nikolaos},
  journal=IEEE_J_CCN,
  volume={6},
  number={2},
  pages={844--857},
  year={2019},
}

@article{badia2024game,
  title={A game of ages for slotted {ALOHA} with capture},
  author={Badia, Leonardo and Zanella, Andrea and Zorzi, Michele},
  journal=IEEE_J_MC,
  year={2024},
volume = 23,
number = 5,
pages = {4878--4889},
}

@article{da2023energy,
  title={Energy-Aware Federated Learning with Distributed User Sampling and Multichannel {ALOHA}},
  author={Da Silva, Rafael Valente and L{\'o}pez, Onel L Alcaraz and Souza, Richard Demo},
  journal=IEEE_J_COML,
  year={2023},
volume= 27, number = 10,
pages = {2867--2871},
}

@article{faddoul2024spatial,
  title={Spatial Modulation With Energy Detection: Diversity Analysis and Experimental Evaluation},
  author={Faddoul, Elio and Kraidy, Ghassan M and Psomas, Constantinos and Chatzinotas, Symeon and Krikidis, Ioannis},
  journal={IEEE Trans. Green Commun. Netw.},
  year={2024},
volume = 8, number = 1, month = mar,
pages = {35--49},
}
 
\end{document}